\documentclass[twocolumn,prl,showpacs,preprintnumbers]{revtex4}
\usepackage{amsmath} 
\usepackage{amsfonts} 
\usepackage{amssymb}
\usepackage{graphicx} 

\newcommand{\half}{\tfrac{1}{2}}

\begin{document}

\preprint{MIT-CTP-3561}
\preprint{gr-qc/0502074}

\title{Relationship between Hawking Radiation and Gravitational Anomalies}
\author{Sean P.~Robinson} 
\email{spatrick@mit.edu}
\author{Frank Wilczek}
\email{wilczek@mit.edu}  
\affiliation{Center for Theoretical Physics, Laboratory for Nuclear
Science and Department of Physics, Massachusetts Institute of Technology,
Cambridge, Massachusetts 02139, USA}

\begin{abstract} We show that in order to
avoid a breakdown of general covariance at the quantum level 
the total flux in each outgoing partial wave of a quantum field in a black hole background must be equal to that of a $(1+1)$ dimensional blackbody at the Hawking temperature.  
\end{abstract}

\maketitle

%%%%%%%%%%%%%%%%%%

{\it Introduction}.---Hawking radiation from black holes is one of the most striking effects that is
 known, or at least widely agreed, to arise from the combination of quantum 
mechanics and general relativity.  Hawking radiation originates upon quantization of matter
in a background spacetime that contains an event horizon---for example, a black 
hole. One finds that the occupation number spectrum of quantum field modes in
the vacuum state is that of a blackbody at a fixed temperature given by the
surface gravity of the horizon. The literature contains several derivations of Hawking radiation, each with strengths and weaknesses.  Hawking's original derivation 
\cite{Hawking:sw, Hawking:rv} is very direct and physical, but it relies on 
hypothetical properties of modes that undergo extreme blue shifts, and specifically assumes that their interactions with matter can be ignored.  Derivations based 
on Euclidean quantum gravity are quick and elegant, but the formalism lacks a 
secure microscopic foundation \cite{Gibbons:1976ue}.  Derivations based on string theory have a 
logically consistent foundation, but they only apply to special solutions in 
unrealistic world-models, and they do not explain the simplicity and 
generality of the results inferred from the other methods \cite{Strominger:1996sh, Peet:2000hn}.   In all these approaches, the Hawking radiation appears as a rather special and isolated phenomenon. 
Here we discuss another approach, which ties its existence to the cancellation of gravitational anomalies.

An anomaly in a quantum field theory is a conflict between a symmetry of the classical action and the procedure of quantization
(see \cite{Bertlmann:xk} for a review). Anomalies in global symmetries can signal new and interesting
physics, as in the original application to neutral pion decay $\pi^0\rightarrow\gamma\gamma$ \cite{Adler:gk, Bell:ts} 
and in 't Hooft's resolution of the $U(1)$ problem of QCD \cite{'t Hooft:fv,'tHooft:up}.   Anomalies in gauge symmetries, however, represent 
a theoretical inconsistency, leading to difficulties with the probability interpretation of quantum mechanics
due to a loss of positivity.    The cancellation of gauge anomalies gives powerful constraints on the charge spectrum of the standard model, which were important historically \cite{smAnom}.  A gravitational anomaly \cite{Alvarez-Gaume:1983ig} is an anomaly in general covariance, taking the form of non-conservation of the energy-momentum tensor.  The simplest case, which will be crucial for us, arises for a chiral scalar field in 1+1 dimensions; the anomaly then reads \cite{Alvarez-Gaume:1983ig, Bertlmann:xk, Bertlmann:2000da}
\begin{equation}
\nabla_\mu T^\mu_\nu= \frac{1}{96\pi\sqrt{-g}}\epsilon^{\beta\delta}\partial_\delta\partial_\alpha\Gamma^\alpha_{\nu\beta}. \label{eq1}
\end{equation}

There are several cases in physics where anomalies have been connected to the existence of current flows.  Pair creation in an electric field has been related to a chiral anomaly \cite{Blaer:1981ps}.
The existence of exotic charges on solitons, with or without the existence of zero modes, has been related to anomalous charge flows that arise in building up the soliton adiabatically \cite{Goldstone:1981kk, callanH}.
Especially closely related to our problem is the connection between anomaly cancellation and the existence of chiral edge states in the quantum Hall effect \cite{girvin}.

Many years ago Christiansen and Fulling \cite{Christensen:jc,Christodoulakis:2001ps} showed that 
it is sometimes possible to use an anomaly in conformal symmetry to derive important constraints on the energy-momentum tensors 
of quantum fields in a black hole background.  This anomaly appears as a contribution to the
trace $T^\alpha_\alpha$ of the energy-momentum tensor in a theory where it vanishes classically. 
By requiring finiteness of the energy-momentum tensor  of massless fields as seen by a freely falling observer at the horizon in $(1+1)$-dimensional Schwarzschild background metric and 
imposing the anomalous trace equation everywhere, one finds an outgoing flux 
given by \( \frac{M}{2}\int_{2M}^{\infty}{\frac{dr}{r^2}T^\mu_\mu(r)}\), where $M$ is the black hole mass, which is in 
quantitative agreement with Hawking's result.  
This is a beautiful 
observation, but it is quite special, and might be regarded an isolated 
curiosity.   Specifically, the limitation to massless fields is quite 
essential to the analysis, as is the limitation to $(1+1)$ dimensions. 
Indeed, only the absence of back-scattering for 
massless particles in $(1+1)$ dimensions allows one to relate flux at the 
horizon---which is the simple, universal aspect of Hawking radiation---to an 
integral over the exterior.   Also, as 
a conceptual matter, the central role ascribed to conformal symmetry seems 
rather artificial in this context.  

 %%%%%%%%%%%%%%%%%%
 
 {\it Framework}.---Our goal is to formulate an effective theory for the behavior of fields in the region
outside the  horizon.  The relevant dynamics of the interior (that is, the part that effects the exterior) is assumed to be captured by an account of the horizon, regarded as a dynamical system.  At the classical level, there is a very useful effective membrane theory of the horizon, which can be derived in a fairly straightforward way from the classical action \cite{Thorne:1986iy, Parikh:1999mf}.   

A delicate issue arises, however, when one moves to the quantum theory.   To identify the ground state of a quantum field (say, for definiteness, a free field), one  normally associates positive energy with occupation of modes of positive frequency.  But in defining positive frequency, one must refer to a specific definition of time.  In the exterior region, where the effective theory is formulated, there is a natural definition of time, for which translation $t \rightarrow t+t_0$ leaves the metric invariant.    This time coordinate becomes mathematically ill-defined at the horizon, and the ``ground state'' associated with its use (Boulware state) is physically problematic, since in it a freely-falling observer would, upon passing through the horizon, feel a singular flux of energy-momentum.  The singular contribution arises from modes that propagate nearly along the horizon at high frequency.
\begin{figure}
\includegraphics[width=246pt]{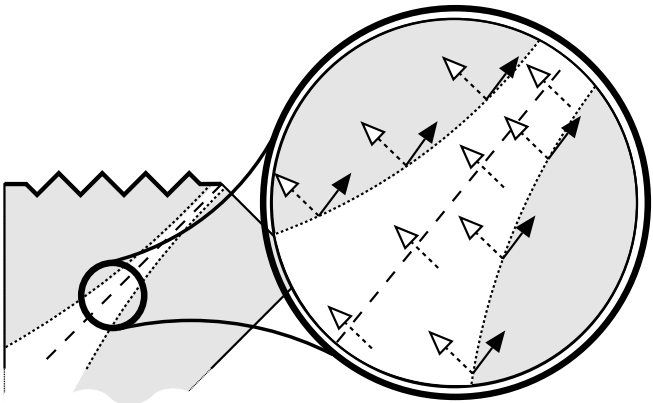}
\caption{\label{fig1} Part of the causal diagram of a black hole spacetime, with inset detail of a region near the horizon. Dashed arrows indicate unoccupied modes, while solid arrows indicate occupied modes. The white region is the infinitesimal slab near the horizon where outgoing modes are eliminated.}
\end{figure}  In the Boulware state, these modes have non-trivial occupation.  The Unruh vacuum \cite{Unruh:db}, which is non-singular, is defined instead by associating positive energy to these modes, so they are unoccupied.   Mathematically, it is implemented by associating positive energy with occupation of modes that are positive frequency with respect to Kruskal $U$.  

Our proposal arises from elevating this state choice to the level of theory choice.  That is, we suppose that the quantum field theory just below the membrane, to which we should join, does not contain the offending modes:  in effect, that they can be integrated out.   

There is an apparent difficulty with this, however.   Having excluded propagation along one lightlike direction, the effective near-horizon quantum field theory becomes chiral.   But chiral theories contain gravitational anomalies, as discussed above.   In our context the 
original underlying theory is generally covariant, 
so failure of the effective theory to reflect this symmetry is a glaring deficiency.  
Analogy to the quantum Hall effect suggests that one might relieve the problem by introducing a compensating real
energy-momentum flux whose divergence cancels the anomaly at the horizon. 
We will show that the energy-momentum associated with Hawking radiation originating at the horizon does the job.
One can extend  the discussion to construct an effective theory for the interior as well as the exterior bulk, separated by a chiral bilayer membrane near the horizon.  In this context, the horizon 
acts as a sort of hot plate, radiating both in to and out of the 
black hole, similar to pair creation in a constant electric field.  

%%%%%%%%%%%%%%%%%%%%%%%%%%%%%%%%

{\it Calculation}.---Consider the partial wave decomposition of
a scalar field in a static, spherically symmetric
background spacetime that is a solution of the 
$d$-dimensional Einstein 
equations with a background matter source such that 
the energy density equals negative the radial pressure.   
In suitable coordinates, the metric of the spacetime can be
written as
\begin{equation}
ds^2 = -f(r) dt^2 + \frac{1}{f(r)} dr^2 +r^2d\Omega^2_{(d-2)}, \label{eq2}
\end{equation}
where $d^2\Omega_{(d-2)}$ is the line element on the $(d-2)$-sphere and $f(r)$
is dependent on the matter distribution. This form encompasses many physically
interesting spacetimes including Schwarzschild, de Sitter, Reissner-Nordstrom, and 
combinations thereof.
An event horizon is a null hypersurface occurring at real, positive, constant coordinate $r=r_H$, where $f(r_H)$ vanishes. We 
consider the case where $f(r)$ has exactly one positive, real root, and the derivatives of $f$ are all finite on the horizon.  
In particular the surface gravity, 
$\kappa\equiv \half(\partial_r f)|_{r_H}$, does not vanish, and $f(r)\rightarrow 2\kappa(r-r_H)$ as $r\rightarrow r_H$.

Upon transforming to the $r_*$ ``tortoise'' coordinate defined by $\frac{\partial r}{\partial r_*}=f(r)$ and performing the partial wave decomposition, one finds 
that the effective radial potentials
for partial wave modes of the scalar field vanish exponentially fast near the horizon. 
Thus physics near
the horizon can be described using an infinite
collection of $(1+1)$-dimensional fields, each propagating in a spacetime with
a metric given by the ``$r-t$'' section of the full spacetime 
metric (\ref{eq2}).   We adopt this simplification.
%In these coordinates, the kinetic terms for the radial wavefunctions---which 
%are labeled by angular momentum quantum numbers---have a simple
%D'Alembertian form. All interaction and effective radial potential terms, 
%however, are multiplied by $f\left(r(r_*)\right)$, which vanishes 
%exponentially near the horizon. Thus, the near-horizon action is that of an 
%infinite collection of free, massless, $(1+1)$-dimensional scalar fields.

For the reasons discussed above, we impose the constraint that outgoing (horizon-skimming) modes vanish near the horizon as a boundary condition.  We take this condition to be localized on a slab of width $2\varepsilon$ straddling the horizon with $\varepsilon\rightarrow 0$ ultimately  (see Fig.~\ref{fig1}). The energy-momentum tensor in this region then 
exhibits an anomaly of the form (\ref{eq1}).

For a metric of the form (\ref{eq2}), the anomaly is purely time-like and 
can be written as 
\begin{equation}
\nabla_\mu {T_\chi}^\mu_\nu \equiv A_\nu \equiv \frac{1}{\sqrt{-g}}\partial_\mu N^\mu_\nu ,
\label{eq3}
\end{equation}
 where  the components of $N^\mu_\nu$ are
\begin{subequations}
\begin{eqnarray}
N^t_t&=&N^r_r=0, \\
N^r_t&=&\frac{1}{192\pi}\left(f^{\prime 2}+f^{\prime\prime}f\right), \\
N^t_r&=&\frac{-1}{192\pi f^2}\left(f^{\prime 2}-f^{\prime\prime}f\right).
\end{eqnarray}
\end{subequations}
%\footnote{
%For reference, we have:
%\begin{subequations}
%\begin{eqnarray}
%\Gamma^t_{tr}=\Gamma^t_{rt}=-\Gamma^r_{rr} &=&\half f^{-1}\partial_r f, \\
%\Gamma^r_{tt}&=& \half f \partial_r f.
%\end{eqnarray}
%\end{subequations}
%All other components vanish.},
%\section{Effective Action}
%
The contribution to effective action for the metric $g_{\mu\nu}$ due to matter fields that interact with this metric is given by
\begin{equation}
W[g_{\mu\nu}]\equiv -i \ln{\left(\int{{\mathcal D}[{\rm matter}]e^{iS[{\rm matter},g_{\mu\nu}]}}\right) },
\end{equation}
where $S[\rm{matter},g_{\mu\nu}]$ is the classical action functional.
Under general coordinate transformations the classical action $S$ changes by
$\delta_\lambda S=-\int{d^dx \sqrt{-g}\lambda^\nu\nabla_\mu T^\mu_\nu}$
where $T^\mu_\nu$ is the energy-momentum tensor and 
$\lambda$ is the variational parameter. 

General covariance of the full quantum theory requires 
$\delta_\lambda W=0$. We write this as
%\begin{widetext}
\begin{eqnarray}
\lefteqn{-\delta_\lambda W 
=\int{d^2x \sqrt{-g}\lambda^\nu\nabla_\mu
\left\{ {T_\chi}^\mu_\nu H  +{T_o}^\mu_\nu\Theta_{+}  + {T_i}^\mu_\nu\Theta_{-}  \right\}} }\nonumber\\ 
&=&\int{d^2x\lambda^t \left\{\partial_r (N^r_t H)
     + \left({T_o}^r_t-{T_\chi}^r_t+N^r_t\right)\partial\Theta_{+} \right.}\nonumber\\
&& \phantom{\int{d^2x\lambda^t \left\{\partial_r (N^r_t H)\right.}}
\left.    + \left({T_i}^r_t-{T_\chi}^r_t+N^r_t\right)\partial\Theta_{-}
   \right\}\nonumber\\
\lefteqn{+\int{d^2x\lambda^r \left\{ 
     \left({T_o}^r_r-{T_\chi}^r_r\right)\partial\Theta_{+} 
     + \left({T_i}^r_r-{T_\chi}^r_r\right)\partial\Theta_{-}
   \right\}}}&&\label{eq6}
\end{eqnarray}
%\end{widetext}
where $\Theta_\pm=\Theta\left(\pm r\mp r_H -\varepsilon\right)$ are scalar step
 functions and $H=1-\Theta_{+}-\Theta_{-}$ is a scalar ``top hat'' function 
which is 1 in the region between $r_H\pm\varepsilon$ and zero elsewhere. 
The anomalous chiral physics is described by ${T_\chi}^\mu_\nu$ via 
Eq.~(\ref{eq3}). The energy-momentum tensors ${T_o}^\mu_\nu$ and ${T_i}^\mu_\nu$ are the covariantly conserved energy-momentum tensors outside and inside 
the horizon, respectively. Constancy in time and Eq.~(\ref{eq3}) together
restrict the form of the ${T}^\mu_\nu$ up to an arbitrary 
function of $r$, which is the trace $T^\alpha_\alpha$, and two constants of
integration, $K$ and $Q$:
\begin{subequations}\label{eq7}
\begin{eqnarray}
T^t_t&=& -(K+Q)/f-B(r)/f-I(r)/f+T^\alpha_\alpha(r),\\
T^r_r&=&  (K+Q)/f+B(r)/f+I(r)/f, \\
T^r_t&=& -K + C(r)  = -f^2 T^t_r, 
\end{eqnarray}
\end{subequations}
where $C(r)=\int_{r_H}^r{A_t(x)dx}$, $B(r)=\int_{r_H}^r{f(x)A_r(x)dx}$, 
and $I(r)=\half\int_{r_H}^r{T^\alpha_\alpha(x)f'(x)dx}$.

A few remarks regarding the evaluation of Eq.~(\ref{eq7}) are in order.
A trace could arise
from a number of physical sources, among them a conformal anomaly. We assume, however, that 
$I/f\Bigr|_{r_H}=\half T^\alpha_\alpha\Bigr|_{r_H}$ is finite.  Since we are concerned with the conditions imposed by canceling potential divergences, finite terms play no role.  Moreover, the terms containing the components
of $A_\nu$ vanish at the horizon.
Note that for the diagonal terms in Eq.~(\ref{eq7}), the limit $r\rightarrow r_H$ depends on whether $r_H$ is approached
from above or below, since 
$f$ flips signs as the horizon is crossed.  The limit from below becomes equal to the limit from above after the  substitution $f\rightarrow -f$.  

We can now take the $\varepsilon\rightarrow 0$ limit of Eq.~(\ref{eq6}). 
The term $\partial_r (N^r_t H)$ vanishes in this limit. 
Using the small $\varepsilon$ expansions
\begin{equation}
\partial_\mu\Theta_\pm = \delta^r_\mu \left[ \pm1-\varepsilon\partial_r\pm\half\varepsilon^2\partial^2_r-\ldots\right]\delta(r-r_H),
\end{equation}
and taking all limits from above,
 the variation (\ref{eq6}) becomes
%\begin{widetext}
\begin{eqnarray}
\lefteqn{\delta_\lambda W =
\int{d^2x\lambda^t \left\{
      \left[K_o-K_i\right]\delta(r-r_H)\right.} }&&\nonumber\\
&&\phantom{+} \left.    -\varepsilon \left[K_o+K_i-2K_\chi-2N^r_t\right]\partial\delta(r-r_H)
   +\ldots\right\}\nonumber\\
&&-\int{d^2x\lambda^r \left\{
     \left[\tfrac{K_o+Q_o+K_i+Q_i-2K_\chi-2Q_\chi }{f}\right]\delta(r-r_H)\right.} \nonumber\\
 &&\phantom{-}\left. -\varepsilon \left[\tfrac{K_o+Q_o-K_i-Q_i}{f}\right]\partial\delta(r-r_H)
   +\ldots\right\}. \label{eq11}
\end{eqnarray}
%\end{widetext}
The ellipses represent higher order terms in $\varepsilon$ with higher 
derivatives of $\delta$-functions; the coefficients of these terms are
simply repetitions of the ones given above. 
The delta functions in Eq.~(\ref{eq11}) indicate that only the on-horizon
values of the energy-momentum tensors will contribute to the possible loss of 
general covariance.
Since $\lambda^t$ and  $\lambda^r$ 
are independent arbitrary variational parameters, each of the four terms
in square brackets above must vanish simultaneously, but
only need do so at $r=r_H$. These four conditions can be solved to give
\begin{subequations}
\begin{eqnarray}
K_o=K_i&=&K_\chi+\Phi,\\
Q_o=Q_i&=&Q_\chi-\Phi,
\end{eqnarray}
\end{subequations}
where
\begin{equation}
\Phi= N^r_t\Bigr|_{r_H}= \frac{\kappa^2}{48\pi}.
\end{equation}
The finite trace terms make no contribution in comparison to the 
divergent $K+Q$ terms. 
These conditions fix the 4 of the 6 constants $Q$ and $K$. The total 
energy-momentum tensor
\begin{equation}
T^\mu_\nu={T_o}^\mu_\nu\Theta_{+} 
                         +{T_i}^\mu_\nu\Theta_{-}
                         +{T_\chi}^\mu_\nu H 
\end{equation}
becomes, in the limit $\varepsilon\rightarrow 0$,
\begin{equation}
T^\mu_\nu={T_c}^\mu_\nu+{T_\Phi}^\mu_\nu,
\end{equation}
where ${T_c}^\mu_\nu$ is the conserved energy-momentum tensor which the
matter in this theory would have without any quantum effects, and
${T_\Phi}^\mu_\nu$ is a conserved tensor with $K=-Q=\Phi$, a pure flux. 

A beam of massless blackbody radiation
moving in the positive $r$ direction at a temperature $T$ has a flux of
the form $\Phi=\frac{\pi}{12}T^2.$
Thus we see that
the flux required to cancel the gravitational anomaly at the horizon has a
form equivalent to blackbody radiation with a temperature given by 
$T=\kappa/(2\pi)$.
This is exactly the Hawking temperature for this spacetime, as could be
determined, for example, by finding the period of Euclidean time required
to remove the conical singularity at the horizon of the Euclideanized metric
 \cite{Gibbons:1976ue}.
Thus, the thermal flux required by black hole 
thermodynamics is capable of canceling the anomaly. 
If we fill each partial wave of the full $d$-dimensional theory so that each one behaves like a $(1+1)$ 
dimensional blackbody source at the Hawking temperature, then we reproduce the core of the standard calculation of black hole emission.  
The actual emission is obtained by propagating the emission from these sources through the effective potential due to spatial curvature outside the horizon.  The resulting radiation observed at infinity is that of a $d$-dimensional
grey body at the Hawking temperature.

%%%%%%%%%%%%%%%%%%%%%%%%

{\it Comments}.---(1) In contrast to the argument immediately preceding, based on gravitational anomaly cancellation, it appears difficult to generalize
the conformal anomaly derivation \cite{Christensen:jc} to arbitrary dimensions using partial wave analysis.  In that framework the connection between the anomaly and the Hawking flux is made
through an integral over all of space. In our framework 
the connection between the anomaly and the Hawking flux is made
through a boundary condition at the horizon, which is accurately described 
using $(1+1)$-dimensional physics, irrespective of the true dimension.   (2) 
Masses, centrifugal barriers, and interactions all drop away near the horizon, 
at least formally (as one sees upon passing to tortoise coordinates).  The 
residual free, $(1+1)$-dimensional wave equations are independent of spin.   
Thus our analysis, which locates the source of Hawking radiation in 
$(1+1)$-dimensional physics near the horizon, is consistent with a simple, 
universal form for that radiation. (3) Comparing the fluxes for thermal 
radiation of massless bosons and fermions in $(1+1)$ dimensions, we find
\begin{equation}
\frac{\int_0^\infty  dk k (e^{k/T} -1)^{-1}}{\int_0^\infty  dk k (e^{k/T} +1)^{-1}} = 2
\end{equation}
This same factor of 2 appears in the relative values of the conformal 
anomalies (central charge) and of the gravitational anomalies.   There does 
not appear to be any comparably simple correspondence in higher dimensions. 
(4) In the context of an eternal black hole, one can find a role for thermal 
radiation {\it incoming\/} to the black hole by imposing additional boundary 
conditions near the horizon of the infinite past ($\iota_-$) that are 
symmetric with the ones we imposed above near the horizon of the infinite 
future.   This corresponds to the Hartle-Hawking state \cite{hh}.  (5)  While 
the arguments advanced here show a pleasing consistency between the existence 
of Hawking radiation flux and gravitational anomaly cancellation, they do not 
in themselves suffice to show that the spectrum of radiation is thermal.  One 
might hope to single out the thermal state by imposing an appropriate symmetry.
   Indeed, thermal states support a form of time-translation symmetry that 
makes sense even near the horizon, namely translation by discrete units $\tau$ 
of imaginary time.  Temperatures of different magnitude can be accommodated as 
different units for the periodicity by $T = 1/\tau$.   If we assume that a 
symmetry of this form exists, then anomaly cancellation fixes the unit.   One 
could certainly wish for a less formal, more physically enlightening 
perspective, however.  

\begin{acknowledgments}
 SPR acknowledges the help of A. Turner, B. Fore, and I. Ellwood.  FW thanks R. Bertlmann and V. P. Nair for informative discussions.  
This work is supported in part by funds provided by the U.S. Department
of Energy (D.O.E.) under cooperative research agreement DE-FC02-94ER40818.
\end{acknowledgments}

\end{document}